\documentclass[prd, twocolumn, nofootinbib]{revtex4}

\usepackage{epsfig} 

\newcommand{\beq}{\begin{equation}}
\newcommand{\eeq}{\end{equation}}
\newcommand{\beqa}{\begin{eqnarray}}
\newcommand{\eeqa}{\end{eqnarray}}
\newcommand{\om}{\Omega_m}
\newcommand{\omt}{\Omega_T}
\newcommand{\weff}{w_{\rm eff}}

\def\fun#1#2{\lower3.6pt\vbox{\baselineskip0pt\lineskip.9pt
  \ialign{$\mathsurround=0pt#1\hfil##\hfil$\crcr#2\crcr\sim\crcr}}}

\begin{document} 

\title{Curved Space or Curved Vacuum?} 
\author {Eric V.\ Linder} 
\affiliation{Physics Division, Lawrence Berkeley Laboratory, 
Berkeley, CA 94720} 

\date{\today} 

\begin{abstract} 
While the simple picture of a spatially flat, matter plus cosmological 
constant universe fits current observation of the accelerated expansion, 
strong consideration has also been given to models with dynamical vacuum 
energy.  We examine the tradeoff of ``curving'' the vacuum but retaining 
spatial flatness, vs.\ curving space but retaining the cosmological 
constant.  These different breakdowns in the simple picture could 
readily be distinguished by combined high accuracy supernovae and 
cosmic microwave background distance measurements.  If we allow the uneasy 
situation of both 
breakdowns, the curvature can still be measured to 1\%, but at the 
price of degrading estimation of the equation of state time variation 
by 60\% or more, unless additional information 
(such as weak lensing data or a tight matter density prior) is included. 
\end{abstract} 


\maketitle 

\section{Introduction} \label{sec.intro}

The acceleration of the cosmic expansion implies that in addition to 
the matter component of our universe there must be another energy density 
component (or a modification of theory that produces an effective 
component).  The first observations of 
acceleration \cite{perl99,riess98} evoked 
this through plots of the amount of this dark energy density 
vs.\ matter density, based on supernova distance-redshift data.  To 
obtain acceleration one needs both a sufficiently negative equation of 
state $w=p/\rho<-1/3$, and enough of this dark energy to overwhelm the 
deceleration due to matter.  The sum of these components could add up to 
the critical density, and hence give spatial flatness, or not.  Similarly, 
while the early plots considered the dark energy to be the 
cosmological constant, $\Lambda$ (with $w=-1$), it had long been 
realized that this was not the only possible accelerating component.  

In looking for concordance between different types of observations, we must 
realize that a point in the cosmological constant density-matter density 
plane, $\Omega_\Lambda$-$\Omega_m$, can actually represent many models, 
and an apparent violation of concordance may simply indicate a breakdown 
in the parameter space.  For example, a confidence contour disjoint 
from the flatness line could be a nonflat, $\Lambda$ universe, or could 
truly be a flat universe, but with dark energy different from the 
cosmological constant.  That is, it could represent a breakdown of 
flatness of space, or of flatness of the vacuum potential energy. 

In this article we analyze distance-redshift relations to distinguish 
these two different classes of breaking the simple cosmological model. 
Section \ref{sec:basic} briefly discusses motivations for spatial curvature 
and dynamical dark 
energy.  In \S\ref{sec:distw} we investigate matching the expansion 
rate and distances in cosmological models as a function of redshift. 
The qualitative difference upon allowing time variation in the dark 
energy equation of state is emphasized in \S\ref{sec:distwa}, and future 
observational constraints simultaneously upon spatial curvature and 
dark energy properties are discussed.  Section \ref{sec:concl} concludes. 

\section{Spatial curvature and vacuum curvature \label{sec:basic}}

Even before the discovery of the accelerating expansion, models with 
spatial curvature or dark energy distinct from a cosmological constant 
were considered.  Indeed, motivated by particle physics, in 1985 the 
cosmological relations for distance, volume, and age tests, and for some 
effects on primordial nucleosynthesis and the cosmic microwave background, 
had been developed for models with 
curvature {\it and\/} dark energy equation of state, by Wagoner (in 
unpublished summer school lectures, with results plotted by Linder; also 
see \cite{lin88a,lin88b}).  In 1986 
a dark energy model with a tilted potential was invented by Linde 
\cite{linde86}, and that year Loh \& Spillar \cite{lohspillar} presented 
the results of observational tests, using galaxy abundances. 

With the success of the inflation picture, spatial curvature came to 
seem repugnant.  However, recent work on inflation 
has renewed interest in the possibility of spatial curvature 
(see \cite{inflacurv,inflacurv2} and references therein).  
While constraints from the cosmic microwave background (CMB) and 
large scale structure show a 
consistency with flatness to 2\% (68\% confidence level \cite{spergel}), 
this assumes dark energy is the cosmological constant, and further 
combines several CMB and large scale structure experiments. 

Flatness of the potential energy in the cosmological constant model of 
dark energy leads to the unhappy property that the energy density is 
unchanging over the history of the universe; hence the value today 
should be the same as in the Planck era.  These eras' energy scales are so 
different, however, that this requires an extraordinary coincidence 
in either initial conditions or the presence of us observing today. 

(In fact, this phrasing of flatness vs.\ curvature in the potential 
is not wholly true.  One can have a ``skating'' model 
with a flat potential but a velocity of the scalar field, $\dot\phi$. 
This will give rise to an equation of state $w'=-3(1-w^2)$, where 
$w'=\dot w/H=dw/d\ln a$.  The dark energy density does evolve, as 
$\rho_{de}\sim (1-w_0)+(1+w_0)a^{-6}$, but this does not avoid fine 
tuning problems (see also \cite{caldlin,liddle,joyce}).  Similarly, the 
linear or tilted potential, while 
not flat, has no curvature, so we have stretched a semantic point in 
the title.) 

The possibility of breakdown of flat space or static vacuum 
should therefore be considered seriously.  
Looking toward future 
precision of observational data, it is worthwhile examining in some 
detail the generalization to non-flat space or non-$\Lambda$ dark energy. 
In the next section we focus on distinguishing the two cases of 
breakdown of the simple model, either with both together 
or one at a time.  \S\ref{sec:distwa} treats the ``double trouble'' case where 
we actually fit for spatial curvature and equation of state properties 
simultaneously. 

\section{Non-flat $\Lambda$ vs.\ Flat non-$\Lambda$ \label{sec:distw}}

The expansion history of the universe is basically 
the Hubble parameter $H(z)=\dot a/a$.  An effective equation of 
state can be defined directly from this, and in particular one can 
consider the effective equation of state due to multiple dark energies, 
or to dark energy plus spatial curvature.  Linder \cite{lingrav} showed 
for an expansion history $(H/H_0)^2=\Omega_m(1+z)^3+\delta H_1^2 
+\delta H_2^2$, where $\delta H_1^2$, $\delta H_2^2$ are two arbitrary 
modifications of the Friedmann equation, that 
\beq 
w_i\equiv -1+\frac{1}{3}\frac{d\ln \delta H_i^2}{d\ln (1+z)}, \label{eq:wh}
\eeq 
and moreover the effective equation of state of the non-matter part of 
the universe is 
\beq 
w_{\rm eff}=w_1\frac{\delta H_1^2}{\delta H_1^2+\delta H_2^2}+
w_2\frac{\delta H_2^2}{\delta H_1^2+\delta H_2^2}\,. 
\label{eq:wwh}
\eeq 
That paper helped pioneer investigation of dark energy models that crossed 
from $w>-1$ to $w<-1$ or vice versa, the so-called phantom divide.  One can 
readily see from Eq.~(\ref{eq:wwh}) that the sum of a model with $w<-1$ 
and a model with $w>-1$ gives just such a crossing. 

For the particular focus of this paper, we can consider 
$\delta H_1^2=\Omega_w e^{3\int d\ln(1+z)\,[1+w]}$, $w_1=w$ 
and $\delta H_2^2=(1-\Omega_T)a^{-2}$, $w_2=-1/3$, where $\Omega_T$ is 
the dimensionless total energy density, matter plus dark energy.  The 
second component then represents spatial curvature, $\omt-1$, which acts like 
a component with $w=-1/3$ in the expansion.  While Eq.~(\ref{eq:wwh}) 
holds for a dark energy with time varying $w$ -- the generic case -- and 
hence we can 
define an effective equation of state with both arbitrary dark energy 
and spatial curvature, for simplicity we here show the formula when 
$w$ is constant: 
\beqa  
w_{\rm eff}(a)=w&[&\Omega_w a^{-3(1+w)}+(-3w)^{-1}(1-\omt)a^{-2}] \nonumber\\ 
&/&[\Omega_w a^{-3(1+w)}+(1-\omt)a^{-2}]. \label{eq:womt}
\eeqa 

\begin{figure}
\begin{center}
\psfig{file=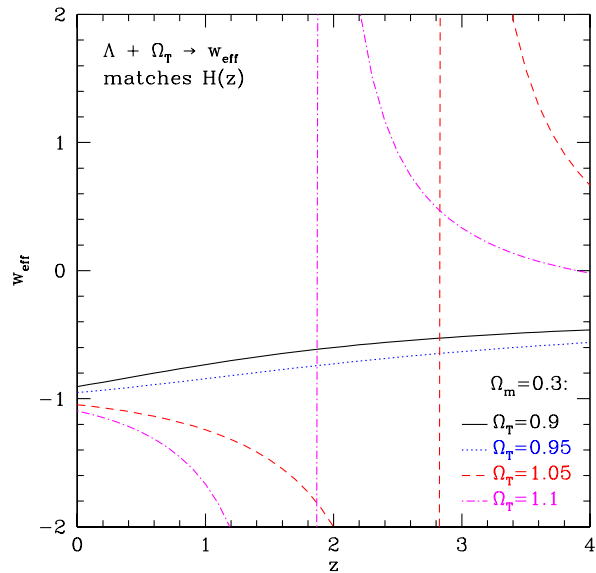,width=3.2in}
\caption{Combining the cosmological constant and spatial curvature, 
$\Omega_T\ne1$, 
leads to an expansion history $H(z)$ of a flat, possibly phantom, 
dark energy model with $w_{\rm eff}(z)$.  Spatial curvature can thus 
be mimicked by dark energy in $H(z)$, but will enter differently in 
distance measurements. 
}
\label{fig:weffh} 
\end{center}
\end{figure}

If the spatial curvature is negative, $\omt<1$, then $w_{\rm eff}$ will 
lie between $w$ and $-1/3$ for all $a$ (even if $w$ is not constant). 
However, if the spatial curvature is positive, $\omt>1$, and dark energy 
dominates curvature then $w_{\rm eff}$ will be more negative than $w$; 
conversely, if curvature dominates dark energy then $w_{\rm eff}$ will 
be more positive than $-1/3$.  The transition, where $\weff$ goes from 
$-\infty$ at $z$ just less than $z_t$, to $+\infty$ just above, occurs at 
$z_t=-1+[(\omt-1)/\Omega_w]^{1/(1+3w)}$.  This behavior is illustrated in 
Fig.~\ref{fig:weffh}.

Note that we readily obtain not only models 
crossing $w=-1$, but can obtain a phantom equation of state, $w<-1$, 
using only components with $w>-1$.  

Few types of observations detect the expansion history independent of 
the geometry of spatial 
curvature, i.e.\ the bare Hubble parameter $H(z)$, with the notable 
exception of the radial modes of baryon acoustic oscillation measurements 
\cite{linbo,gb03,seoeis03}.  Far more common are distance measurements, 
involving not only an integral of the Hubble parameter but a functional 
dependence on the spatial curvature.  One can write the comoving distance as 
\beq 
d(z)=(1-\omt)^{-1/2}\sinh[(1-\omt)^{1/2}\int_0^z dz'/H(z')], \label{eq:dcurv} 
\eeq 
remembering that $\sinh$ is a complete function (i.e.\ the expression is 
valid for all $\omt$). 

Now let us consider how models with spatial curvature, but simple $\Lambda$, 
can mock up universes with different dark energy but retaining spatial 
flatness, or vice versa.  To focus on the tradeoff between non-flat space and 
dynamic vacuum, we only consider these one at a time for the rest of this 
section.  That is, we want to know when the distance to some redshift 
$z$ in a $\Lambda$ universe but with spatial curvature matches the 
distance to the same redshift in a spatially flat universe with constant $w$ 
(initially, for simplicity). 
So we set $d(\om,\omt,w=-1;z)=d(\om',\omt=1,w;z)$ for some $z$.  
We take $\om'=\om$ to concentrate on the effects of $\Omega_T$ and $w$ 
(also see \S\ref{sec:matchwofz}). 

\begin{figure}
\begin{center}
\psfig{file=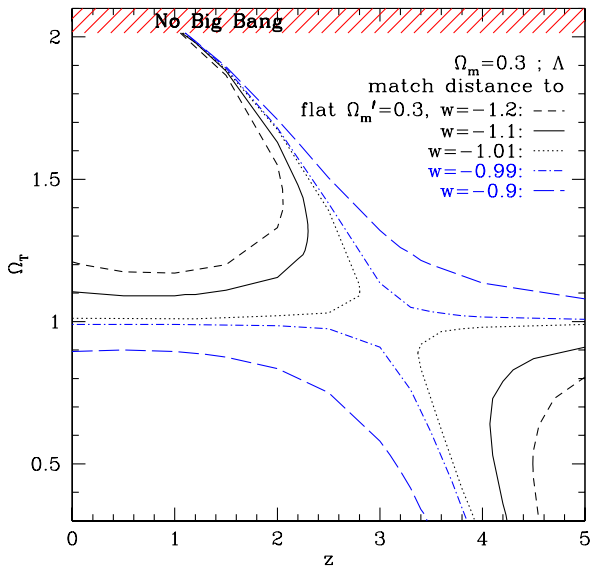,width=3.2in}
\caption{Matching distances in a non-flat, $\Lambda$ model 
and a flat, dark energy model indicates the interplay between the 
total energy density $\Omega_T$ needed for each redshift ($\omt$ of 
course is constant in any one model) 
and the dark energy equation of 
state $w$.  Both cases keep $\om=0.3$.  This interplay varies greatly 
over redshift (see the text for detailed discussion), pointing out 
the need for distance observations over a range of redshifts to distinguish 
curvature from dark energy.  These can be provided by, e.g., supernova 
measurements to $z\lesssim2$ plus CMB measurements to $z=1089$. 
}
\label{fig:omtz} 
\end{center}
\end{figure}

We find a complex, many-to-one, relationship between the spatial curvature 
in terms of $\omt$, the dark energy equation of state $w$, and the redshift 
of the observations.  Figure~\ref{fig:omtz} shows the spatial curvature 
needed at each redshift in a $\Lambda$ model to match distance observations 
to flat universes with given values of $w$.  

At low redshift, 
there is a one-to-one mapping (basically given by the deceleration 
parameter $q_0$) between $\omt$ and $w$.  In the $\omt>1$ case, 
$\Omega_\Lambda$ has been increased so in the flat model, with 
$\Omega_w$ fixed to $1-\om$, $w$ 
must become more negative to obtain the same acceleration.  However at 
high redshift the positive curvature becomes important, decreasing the 
distance to a given redshift, and so is matched by having less 
acceleration or a less negative $w$.  Thus the matching model must 
cross from $w<-1$ at low redshift to $w>-1$ at high redshift.  The reverse 
occurs for $\omt<1$.  At some 
point in between of course the two effects in the $\omt\ne1$ model 
cancel out and the distance is insensitive to the spatial curvature. 
The redshift dependence of interpreting $\omt$ vs.\ $w$ in distance 
measurements was pointed out in \cite{lin01}, and the nulling 
redshift explicitly discussed in \cite{linbo5}. 

In addition, at redshifts above and below the nulling redshift the 
matching becomes double-valued.  For example a flat model with $w=-1.1$ 
matches the distance to $z=1.5$ of a $\Lambda$ model with either 
$\omt=1.11$ or 1.88.  Likewise a flat model with $w=-0.9$ matches a 
$\Lambda$ model with $\omt=0.875$ or 1.895.  This double valued behavior 
is cut off at high $\omt$ (and low redshift) by requiring a Big Bang; 
larger values of $\Omega_\Lambda$ would cause a bounce in the past.  Note 
that there is an attractor on the upper branch, where different values 
of $w$ match the same $\omt$ as they approach the ``no Big Bang'' state.  
At low $\omt$ the cut off is due to requiring positive $\Lambda$.  At 
asymptotically high $z$, there is an attractor at $\omt=1$, since in 
the flat model the dark energy fades away and the expansion sees a pure 
matter dominated state. 

The implications of this distance matching are: 1) At low redshifts, 
$z\lesssim1$, curvature is roughly degenerate with a shift in $w$ from 
$-1$.  Distance observations at somewhat higher redshifts can break this.  2) 
Distance observations at the nulling redshift ($z=3.1$ for a fiducial 
flat, $\om=0.3$ model) are quite insensitive to curvature.  This is a 
favored redshift range for ground based baryon acoustic oscillation 
measurements, meaning that such alone will not be able to provide tight 
constraints on spatial curvature. On the plus side, at this nulling redshift 
the variance of other parameters will be lessened because of zero 
covariance with curvature.  3) Distance measurements at very high redshift 
again have some sensitivity to curvature, and as this redshift increases 
one gets the best of points 1 and 2: the curvature influences the 
distance, but has reduced degeneracy with the dark energy properties. 

A nice extension of \cite{lin01} quantifying the tradeoff of $\omt$ vs.\ $w$ 
is given by Polarski \& Ranquet \cite{polarski}.  Their Fig.~1, valid for 
$z=1.5$ (see also their Fig.~3 which has some $z=4$ information), clearly 
shows some of the degeneracies discussed.   To see the progression of 
degeneracies and covariances with redshift, see Fig.~\ref{fig:omtwflower} 
below.  For illustration, this has been highly idealized, with other 
parameters fixed and 
distance measurements of 0.02\%, constant in redshift.  As expected, we 
see the degeneracy direction is the same at $z\approx0-1$, agreeing with that 
of $q_0$; at the nulling redshift $z=3.1$ the parameters are decorrelated 
and $w$ is determined much better than $\omt$; high redshift measurements 
are orthogonal to low redshift ones, and asymptotically the parameters 
decorrelate again, this time in favor of $\omt$.

\begin{figure}
\begin{center}
\psfig{file=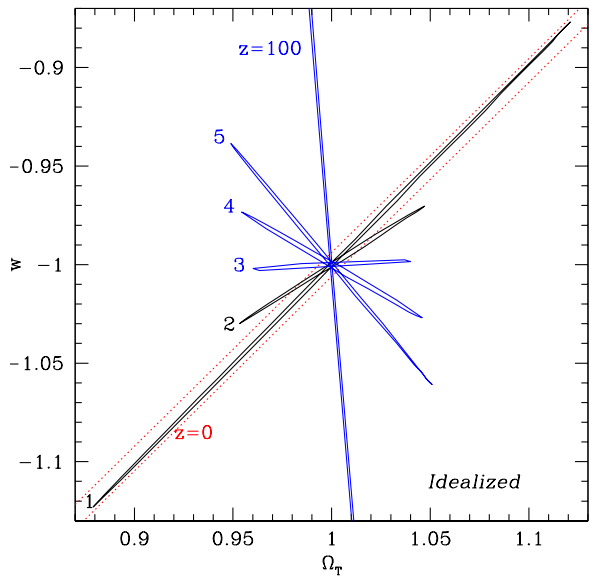,width=3.2in}
\caption{Covariances between total energy density $\omt$ and constant 
equation of state $w$ are illustrated for idealized distance data at 
various redshifts (as labeled).  Other parameters are fixed.  The 
nulling redshift 
for curvature information is at $z\approx3$ and high redshift distances 
(such as to the CMB last scattering surface) are more sensitive to 
curvature than to $w$. 
}
\label{fig:omtwflower} 
\end{center}
\end{figure}

Several immediate cautions are in order.  This diagram was highly idealized 
to bring out the degeneracy behavior.  Realistic statistics, systematics 
(in particular gravitational lensing for high redshift measurements 
\cite{holzlin}), 
and marginalizing over the other parameters, such as the matter density 
and absolute source luminosity, will greatly affect the measurements and 
their orthogonality at different redshifts.  As one example, see \cite{linsl} 
where the orthogonality in the case of a strong lensing probe is completely 
undone on consideration of the realistic case.  Furthermore,  
constant equation of state is a very specialized assumption and does not 
give the generic degeneracy behavior of models with time varying equation 
of state.  We address these issues later with Fisher analysis. 

Another issue is the dependence on the fiducial value of the matter 
density $\om$.  For a given value of $w$ and redshift, the matching 
$\omt$ changes 
by 1\% on varying $\om$ between 0.2 and 0.4, except at higher redshifts 
or on the anomalous branch it can change by $\sim0.2$.  For a given 
value of $\omt$ and redshift, the matching $w$ changes by about 1\% over 
the same $\om$ range, or up to 2.5\% at high $z$. 

\begin{figure}
\begin{center}
\psfig{file=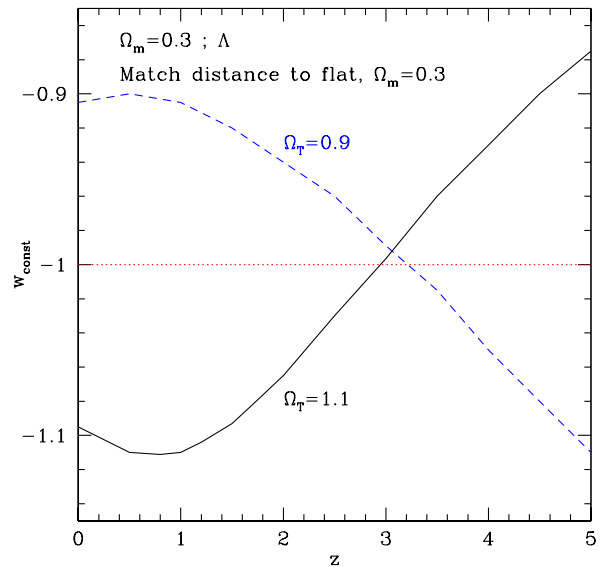,width=3.2in}
\caption{Matching distances in a non-flat, $\Lambda$ model 
and a flat, dark energy model allows derivation of the constant 
equation of state value, $w_{\rm const}$, that fits at any one 
redshift.  The variation of these 
values from a constant indicates how strong the degeneracy is between 
curvature and constant $w$.  Both models keep $\om=0.3$.  
}
\label{fig:matchwz} 
\end{center}
\end{figure}

Finally, we can ask conversely what value of constant $w$ would be needed 
in a flat universe to match distance measurements from a nonflat, $\Lambda$ 
universe, as a function of redshift.  This is shown in Fig.~\ref{fig:matchwz}. 
Note that this is {\it not\/} an evolving $w(z)$ but rather a different 
constant $w$ to match the distance to each separate redshift.  It merely 
provides a measure of how 
discrepant a single constant $w$ would be, just as in Fig.~\ref{fig:omtz} 
the spatial curvature was not evolving but we wanted to see how the 
{\it derived\/} matching $\omt$ from each redshift would look.

Again we see that at low redshift, $z<1$, a constant $w$ is a reasonable 
fit to match spatial curvature.  At higher redshifts this degeneracy is 
broken.  The nulling near $z\approx3$ where the curves cross is clear, 
where very different values of spatial curvature, positive or negative, 
match.  In the next section we extend the analysis to models with time 
varying equation of state, as this figure clearly indicates we must.

\section{Fitting Non-flat Space and Non-static Vacuum \label{sec:distwa}} 

An orthogonal question to those we have asked so far is what if the 
measurements imply the simple flat, $\Lambda$ model.  While simplicity 
would then argue for this being the favored solution, one could consider 
the ``double trouble'' case of allowing both non-flatness of space and 
dynamics in the vacuum to mock up the simple model.  Through Fisher 
analysis we 
can see the degeneracies that would allow this and estimate the parameter 
constraints in a fit simultaneously accounting for $\omt$ and $w$. 

\subsection{Fitting $w_{\rm const}$ and $\Omega_T$ \label{sec:womt}} 

From the Fisher sensitivities of the parameters, $\partial 
d/\partial{\rm par}$, we can derive trade-off rules for one parameter to mock 
up the effects of another.  For a fiducial model near the flat, $\om=0.3$ 
plus $\Lambda$ model,  the trade is 
\beqa 
\Delta\omt&\approx& \Delta w, \quad (z\ll1) \\ 
&\approx& 17\Delta w, \quad (z=3) \\ 
&\approx& -0.06\Delta w, \quad (z\approx z_{lss}), 
\eeqa 
where $z_{lss}=1089$ is the redshift of the CMB last 
scattering surface. 
These relations exhibit the sign change from low to high redshift and 
change in sensitivity discussed earlier.  CMB results are seen to be 
more sensitive to spatial curvature than to the value of $w$, as is 
well known (e.g.\ \cite{hu}).  

One also needs to take account of the covariances among the entire 
parameter set.  For supernova (SN) distance measurements one has a parameter 
set of $\{{\cal M},\om,\omt,w\}$, where ${\cal M}$ is a nuisance parameter 
combining the supernova absolute luminosity and the Hubble constant. 
Relaxing spatial flatness has a significant effect on parameter estimation. 
Even with constant $w$, data strongly constraining ${\cal M}$ (e.g.\ 
low redshift SN such as from the Nearby Supernova Factory \cite{snf}) 
and a tight prior on $\om$ (of order 0.01) are required for reasonable 
estimations. 
With both of these, and distance data from $z=0-1.7$ of the quality 
expected from SNAP \cite{snap}, the marginalized errors are 
$\sigma(\omt)=0.19$ and $\sigma(w_{\rm const})=0.20$.  Contrast this with 
the constraints $\sigma(\omt)=0.065$ when we fix $w=-1$ or $\sigma(w_{\rm 
const})=0.065$ when we fix $\omt=1$ -- without an $\om$ prior.  The 
covariances greatly increase uncertainties over na{\"\i}ve expectations. 

Complementarity with high redshift distances ($z\gtrsim4$) is expected 
to be useful, but these distances must be known with strong statistical and 
systematic accuracy -- a challenging task.  However we found in 
\S\ref{sec:distw} that distances to $z\gg1$ would be most useful, and the 
CMB data provides accurate distance measurements to $z=1089$.  Adding CMB 
data of 0.7\% precision in the reduced distance to last scattering and 
0.9\% on $\om h^2$ (as from the next generation Planck Surveyor 
experiment \cite{planck}) to SN data, and including marginalization over the 
reduced Hubble constant $h$, yields a simultaneous estimation with 
$\sigma(\omt)=0.011$ and $\sigma(w_{\rm const})=0.079$ without any 
additional priors.  By contrast, adding 
even a very optimistic 1\% measurement of the distance to $z=5$, which 
in Fig.~\ref{fig:omtwflower} looks so orthogonal to $z\approx1$ measurements, 
only gives $\sigma(\omt)=0.075$ and $\sigma(w_{\rm const})=0.19$.  Thus, 
SN+CMB data provide good constraints in the ``double trouble'' scenario 
of breaking spatial flatness and vacuum staticness. 

\subsection{Matching Time Varying Equation of State \label{sec:matchwofz}} 

Note that while we have marginalized over the matter density parameter, 
we have required that $\om$ in the flat model be the same as the $\om$ in the 
nonflat matching model, so far in this paper.  The reason is that at 
high redshift matching the distances in the matter dominated era 
requires that $w\to0$ to offset the difference between models with 
different $\om$ (cf.\ Eq.~\ref{eq:wh}).  However, since acceleration 
today implies $w<-1/3$ 
then matching between models with different $\om$ cannot be accomplished 
with constant $w$.  Due to this, and the results of Fig.~\ref{fig:matchwz}, 
we now consider models with spatial curvature and 
time varying equation of state. 

As a first step, let us go back and look at matching the expansion 
history $H(z)$.  As stated in the beginning of \S\ref{sec:distw}, 
Eq.~(\ref{eq:wwh}) holds for time varying $w$ and Eq.~(\ref{eq:womt}) 
can be easily generalized by changing $a^{-3(1+w)}$ to $exp\{3\int_a^1 
d\ln a\,[1+w(z)]\}$.  Matching the expansion $H$ is not appropriate for 
measurements of distance, because of the extra impact of spatial curvature 
as seen in Eq.~(\ref{eq:dcurv}).  Suppose we wanted to match the 
distance in a nonflat, $\Lambda$ model to that in a flat, $w(z)\ne-1$ 
model, at every redshift.  Upon differentiating the respective distances 
with respect to redshift, we find the requirement 
\beqa 
H(w,\omt=1,\om';z)&=&H(w=-1,\omt,\om;z) \label{eq:hwhlam} \\ 
&\quad& /\cosh\left(\sqrt{1-\omt}\int_0^z dz'/H_\Lambda\right). \nonumber
\eeqa 
Clearly, matching $H$ will not lead to matching distances, and vice versa. 

In fact, in general one cannot derive a well behaved $w(z)$ that matches 
the behavior of a nonflat, $\Lambda$ model.  Recall that 
Fig.~\ref{fig:matchwz} showed a sequence of constant $w$ models, where 
the distance to redshift $z$ was given in terms of $w$ constant from 0 
to $z$, and not a true $w(z)$.  One might think that one could form a 
function $w(z)$  by interpolating between the constant $w$'s but this 
is fruitless.  

An analogy is attempting to match the position of a 
car on a highway at all times.  Say that a car behind our car 
is traveling at a constant 80 kph and a car ahead of our car 
travels at a constant 100 kph.  Our car's behavior is not bounded between 
80 and 100 kph; it can go at 120 kph, and then 50 kph, say, and its 
position will still remain between the other two cars.  Mathematically, 
the mean value theorem does not guarantee that derivatives of the 
function are bounded within the derivatives of the bounding functions.  

If one attempts to derive a function $w(z)$ by defining it in terms of 
the derivative of the Hubble parameter, i.e.\ Eq.~(\ref{eq:wh}) applied 
to Eq.~(\ref{eq:hwhlam}), then one runs into pathologies.  This is due 
to the cosh factor which reduces $H_\Lambda$.  Eventually the cosh grows 
large enough that the right hand side is driven below $\om (1+z)^3$, 
effectively requiring a negative energy density for the dark energy, 
and no value of $w$ can handle the crossover to negative energies. 
Instead, one must be content with a derived $H(z)$, given by 
Eq.~(\ref{eq:hwhlam}).  For $\omt>1$, and $\om'\le\om$, one can define 
a $w(z)$, though its behavior is somewhat baroque. 

At low redshift, however, there is no problem in deriving the 
distance-matching $w$.  To lowest order in $z$, this is given as before 
by matching the deceleration parameter $q_0$.  Then 
\beq 
w_0=-\frac{1}{3}+\frac{1}{3}\frac{1}{1-\om'}[\om-\om'-2(\omt-\om)], 
\label{eq:wq0} 
\eeq 
where we can now allow the matter density $\om'$ of the flat, $w\ne-1$ model 
to differ from $\om$ of the nonflat, $\Lambda$ model. 
As discussed above, there is no point in deriving higher order terms 
in $w(z)$ because we know they will fail as $w$ blows up.  However, we 
are primarily interested in the measured quantities, distances, rather 
than $w$ in this application.  We can find the distance given by this 
$w_0$ and, although it cannot provide a perfect match to the nonflat 
model, ask how good an approximation it is.  

\begin{figure}
\begin{center}
\psfig{file=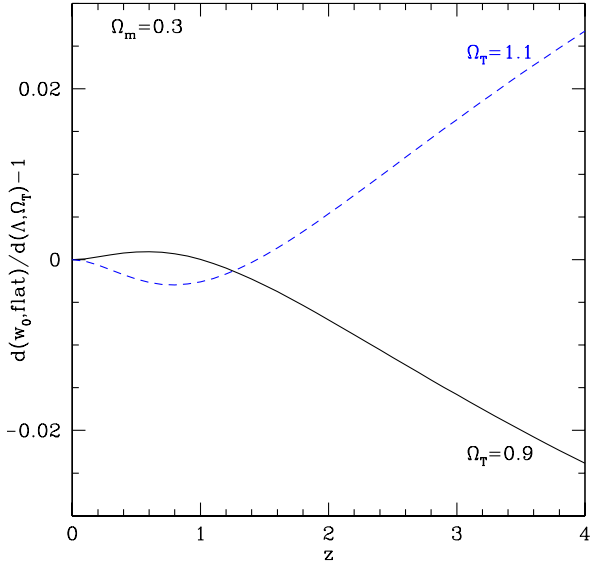,width=3.2in}
\caption{Matching distances at several redshifts in a nonflat, 
$\Lambda$ model and a flat, dark energy model cannot be done with an 
equation of state constant with redshift.  However, the fractional 
distance errors between these cases, shown here, are 
not severe out to $z\approx2$ if the constant $w$ is chosen so as to match 
the deceleration parameters today (i.e.\ give matching distances at 
low redshift).  For $\omt=0.9$ (1.1), this means $w_0=-0.905$ (-1.095). 
Thus curvature could be confused with dark energy possessing constant 
equation of state, but not general, varying $w(z)$.  Combining $z<2$ 
distances with the CMB last scattering distance can break the constant $w$ 
degeneracy. 
}
\label{fig:wlamdist} 
\end{center}
\end{figure}

Figure~\ref{fig:wlamdist} illustrates that in fact it works well. 
Even though the derived $w_0$ is only valid for low redshift, this 
model approximates the nonflat, $\Lambda$ distances to better than 0.5\%, 
or 0.01 mag, out to $z=1.8$.

\subsection{Fitting $w_0$, $w_a$, and $\omt$ \label{sec:wzomt}} 

Finally, we consider the ``triple trouble'' scenario.  We can 
examine future distance data and ask how stringent the simultaneous 
constraints on all the parameters are, including spatial curvature 
through $\omt$ and vacuum dynamics through the dark energy equation of 
state characteristics $w_0$ and $w_a$, with $w(a)=w_0+w_a(1-a)$. 
Again, the combination of SN+CMB data is crucial to provide reasonable 
constraints.  Furthermore, because of the extra degree of freedom in 
$w(a)$, without additional constraints the errors on the equation of 
state blow up, especially on the time variation $w_a$.  
The uncertainties in parameter 
estimation are $\sigma(\omt,w_0,w_a)=0.047$, 0.12, 1.75.

\begin{figure}
\begin{center}
\psfig{file=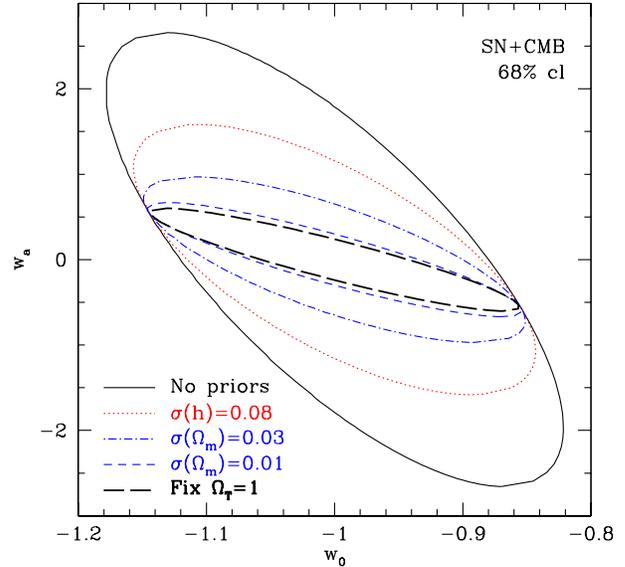,width=3.2in}
\caption{Allowing ``triple trouble'' of a non-flat, non-$\Lambda$, 
time varying equation of state universe strongly increases the 
uncertainties in the cosmological parameters, even for the combination 
of future supernovae and CMB data.  Additional information, e.g.\ 
constraining the matter density through weak gravitational lensing 
data, could break the degeneracy.  Combining all probes, however, 
removes the important element of crosschecks. 
}
\label{fig:curvwa} 
\end{center}
\end{figure}

Figure \ref{fig:curvwa} illustrates the severe degradation upon simultaneously 
estimating $\omt$, $w_0$, and $w_a$, in contrast to assuming spatial 
flatness.  Other parameters are marginalized over. To retain leverage on 
dark energy, an additional constraint is needed, for example in the form of 
a prior on the matter density, or other data such as weak gravitational 
lensing information.  
With a 0.03 
prior on $\om$, the constraints improve to 0.014, 0.098, 0.64.  A 
prior of 0.01 is required to bring the degradation in uncertainties 
below 10\% relative to the fixed spatial flatness case, which has 
$\sigma(w_0)=0.095$, $\sigma(w_a)=0.39$.  Combining with 
weak lensing or other large scale structure data would serve well 
instead, but then no room is left for crosschecks.   Alternatively, 
the matter density can be constrained through the CMB -- if the 
Hubble constant $h$ is accurately determined, as Hu \cite{hutucson} 
has clearly emphasized.

\section{Conclusion \label{sec:concl}} 

The exciting and fundamental property of the accelerating expansion of 
our universe is currently consistent with an explanation in terms of 
a simple cosmology flat in space and static in vacuum energy.  Breakdown of 
one or the other of these properties can be discerned, and distinguished, 
with accurate distance measurements over a wide range of redshifts. 
Distances to the nulling 
redshift $z\approx3$ are fairly insensitive to spatial curvature, and 
measurements at $z>3$ face severe statistical and systematic challenges  
-- except for the CMB. 
For both theoretical and practical reasons, the best tools for the 
purpose of learning about curved space and dynamic vacuum are the 
combination of supernovae and cosmic microwave background distance data.  

Other probes could contribute, such as baryon acoustic oscillations 
measurements of reduced distances and the reduced Hubble parameter, though 
in the $z\approx3$ 
region this will mostly be in the form of breaking parameter degeneracies 
because of the null in sensitivity to spatial curvature.  
We have not considered the role of weak lensing or other large scale structure 
data here, though it can be quite helpful in constraining curvature 
\cite{bernstein,snap,knox}, since if all probes are combined together 
then the important ability to crosscheck the results vanishes. 

While a nonflat, $\Lambda$ universe can be distinguished from a flat, 
$w\ne-1$ universe, and SN+CMB data offer 
good constraints in the ``double trouble'' scenario
of breaking both spatial flatness and vacuum staticness with a constant dark 
energy equation of state, this treatment of dark energy is highly 
nongeneric.   Once we allow triple trouble -- the possibility that 
$\omt\ne1$, $w\ne-1$, and $w\ne$ const (e.g.\ $w_a\ne0$) -- there is 
indeed trouble.  With a 
realistic treatment of dark energy in terms of a time varying equation 
of state, removing the theoretical prejudice for spatial flatness greatly 
weakens our ability to probe the cosmology, even with next generation 
measurements.   

Since in this case the dimensionless 
total energy density of the universe, the dark energy density, and the 
dark energy equation of state all evolve, theoretical prejudice is 
actually also motivated by multiple coincidence problems -- why are 
we living at a time when we can see a nearly flat universe, one with 
roughly equal densities in dark energy and matter, and one accelerating 
with a behavior close to that from a cosmological constant.  Perhaps 
as a first step we can use 
Occam's razor, that entities should not be multiplied beyond necessity, 
backed up by inflation and the instability of $\omt\ne1$, 
to take spatial flatness as an assumption until contradicted by data.  
More satisfactory would be precision determination of $\Omega_m$ or $h$, 
to provide consistent constraints breaking the degeneracies.

\begin{acknowledgments} 

This work has been supported in part by the Director, Office of Science,
Department of Energy under grant DE-AC02-05CH11231.  I thank Andy 
Albrecht, Wayne Hu, Nemanja Kaloper, Alex Kim, 
and Ramon Miquel for useful conversations, and especially Natasha Cayco-Gajic 
and the LBL High School Summer Research Participation Program for motivating 
the search for maximum simplicity in explaining the role of curvature. 

\end{acknowledgments}

\end{document}